\documentclass[preprint,showpacs,preprintnumbers,amsmath,amssymb]{revtex4}
\usepackage{graphicx}
\usepackage{dcolumn}
\usepackage{bm}

\font\scripti=cmmi7
\font\scriptscripti=cmmi5
\def\sib#1{\setbox0 = \hbox{\scripti #1}
  \kern-.02em\copy0\kern-\wd0
  \kern.04em\box0} 
\def\ssib#1{\setbox0 = \hbox{\scriptscripti #1}
  \kern-.02em\copy0\kern-\wd0
  \kern.04em\box0} 
\font\tenib=cmmib10 
\skewchar\tenib='177 \skewchar\tenib='177 \skewchar\tenib='177
\def\pbold#1{\setbox0 = \hbox{$ #1 $}
  \kern-.022em\copy0\kern-\wd0
  \kern.011em\copy0\kern-\wd0
  \kern.011em\copy0\kern-\wd0
  \kern.011em\copy0\kern-\wd0
  \kern.011em\box0} 

\usepackage{graphicx}
\usepackage{dcolumn}
\usepackage{bm}

\def\up{\uparrow}
\def\dwn{\downarrow}

\def\lesssim{\ \raise.3ex\hbox{$<$}\kern-0.8em\lower.7ex\hbox{$\sim$}\ }
\def\gesim{\ \raise.3ex\hbox{$>$}\kern-0.8em\lower.7ex\hbox{$\sim$}\ }

\begin{document}
\title{Precursor of Superfluidity in a Strongly Interacting Fermi Gas \\ with Negative Effective Range}
\author{Hiroyuki Tajima}
\affiliation{Theoretical Research Division, Nishina Center, RIKEN, Wako, Saitama, 351-0198, Japan}
\date{\today}
\begin{abstract}
We theoretically investigate the effects of pairing fluctuations in an ultracold Fermi gas near a Feshbach resonance with a negative effective range.
By employing a many-body $T$-matrix theory with a coupled boson-fermion model,
we show that the single-particle density of states exhibits the so-called pseudogap phenomenon which is a precursor of superfluidity induced by strong pairing fluctuations.
We clarify the region where strong pairing fluctuations play a crucial role in single-particle properties, from the broad-resonance region to the narrow-resonance limit at the divergent two-body scattering length.
We also extrapolate the effects of pairing fluctuations to the positive-effective-range region from our results near the narrow Feshbach resonance.
Results shown in this paper are relevant to the connection between ultracold Fermi gases and low-density neutron matter from the viewpoint of finite-effective-range corrections.
\end{abstract}
\maketitle
\par
\section{Introduction}
The realization of superfluidity in ultracold Fermi gases with pairing interactions, that are tunable by Feshbach resonances, is one of the most important breakthroughs in condensed matter physics \cite{Regal,Zwierlein,Giorgini,Bloch,Chin}.
The Bardeen-Cooper-Schrieffer-Bose-Einstein-condensation (BCS-BEC) crossover phenomenon \cite{Eagles,Leggett,Nozieres,SadeMelo} realized in $^6$Li \cite{Regal} and $^{40}$K \cite{Zwierlein} Fermi gases has been extensively discussed in various fields such as FeSe superconductors \cite{Kasahara1,Kasahara2,Yang}, electron-hole systems \cite{Tomio,Zenker}, nuclear matter \cite{Lombardo,Matsuo,Margueron,Mjin}, and color superconductivity in high-density quark matter \cite{Abuki2,Nishida,Abuki1}. 
\par
In particular, the similarity between ultracold Fermi gases and dilute neutron matter  in a neutron star has recently gathered much attention \cite{Gezerlis,Forbes,Pieter1,Schwenk}.
The idea is based on the fact that both systems are dominated by low-energy $s$-wave scatterings and the temperature $T$ is very low compared to the Fermi temperature $T_{\rm F}$.
Since the neutron-neutron scattering length $a_{\rm nn}=-18.5 {\rm fm}$ \cite{Wiringa} is negatively large and the dimensionless interaction parameter is typically given by $1/k_{\rm F}a_{\rm nn}\simeq-0.03$ (where $k_{\rm F}$ is the Fermi momentum) at the nuclear saturation density $\rho_0\simeq0.17 {\rm fm}^{-3}$, the system property is very close to a unitary Fermi gas ($1/k_{\rm F}a=0$, where $a$ is the two-body scattering length of Fermi atoms). 
In this regard, ground-state thermodynamic quantities have been experimentally measured with high precision \cite{Tajima4,Horikoshi}.   
Moreover, the pairing gap \cite{Chin2,Schirotzek,Hoinka}, critical temperature \cite{Luo,Inada,Ku}, and thermodynamic quantities at finite temperature \cite{Ku,Horikoshi2,Nascimbene2}, which are important information for the cooling mechanism \cite{Flowers,Page2,Shternin,Page,Oertel,Sedrakian} as well as glitch phenomena \cite{Anderson,Delsate,Ho} in a neutron star, have been also measured near the unitarity limit.
\par
However, in addition to the scattering length, there is another key parameter, that is, the effective range $r_{\rm e}$. 
While BCS-BEC crossover physics in ultracold Fermi gases are usually discussed with the zero-range contact-type interaction because the effective range is negligible near the broad Feshbach resonance, the effective range of neutron-neutron scatterings $r_{\rm e,nn}=2.8 {\rm fm}$ \cite{Wiringa} is not negligible in the relevant density region of a neutron star. 
In this regard, effective-range corrections should be considered if one uses to treat an ultracold Fermi gas as a quantum simulator of neutron star matter. 
There are some theoretical studies on these corrections at $T=0$ based on Quantum Monte-Carlo simulations \cite{Gezerlis,Forbes,Schonenberg} and the effective-range dependence of the ground-state energy has been reported. 
On the other hand, although its sign is generally different, finite {\it negative} effective range can be realized in an ultracold Fermi gas with narrow Feshbach resonances \cite{Hazlett}.
We note that recently the optical control of scattering parameters with magnetic Feshbach resonance has also been proposed \cite{Bauer,Wu,Wu2} and experimentally examined in a $^6$Li Fermi gas \cite{Semczuk,Jagannathan}. 
\par
In this paper, we show how negative-effective-range corrections affect system properties in the presence of strong pairing fluctuations near the superfluid phase transition temperature $T_{\rm c}$.
It is well-known that a precursor of the superfluid phase transition can be seen in a strongly interacting Fermi gas through various physical quantities, (e. g., the enhancement of specific heat \cite{Ku,Pieter2} and suppression of spin susceptibility \cite{Wlazlowski,Kashimura,Palestini, Mink,Enss,Tajima,Tajima2,Tajima3,Jensen}).
These strong-coupling effects are deeply related to the so-called pseudogap phenomenon \cite{Tsuchiya,Tsuchiya2,Tsuchiya3,Watanabe,Magierski,Su,Mueller,Mueller2}, where the single-particle density of states near Fermi level shows a dip structure even above $T_{\rm c}$.
Although the pseudogap in an ultracold Fermi gas has not been directly observed in the experiment yet (indirectly observed in photo-emission spectra \cite{Stewart,Gaebler,Sagi2}), it exhibits when and how the Cooper pairing occurs from the microscopic viewpoint when the temperature approaches $T_{\rm c}$ in the normal phase.  
One can expect that such pairing properties have an important role in the cooling process of a neutron star across $T_{\rm c}$.
Actually, the pseudogap phenomenon has been also discussed in dilute nuclear matter \cite{Schnell,Lee2,Abe}.
\par
We numerically calculate the single-particle density of states in a strongly interacting Fermi gas with negative effective range within the framework of the non-selfconsistent $T$-matrix approximation, which have been extensively used for the study of pseudogap physics in this system \cite{Tsuchiya,Watanabe,Mueller}.
To reproduce the finite negative effective range associated with the narrow Feshbach resonance, we employ the so-called coupled fermion-boson model \cite{Timmermans,Holland,Ohashi2,Liu}.
We obtain the pseudogap temperature $T_{\rm pg}$, which is a characteristic temperature where pairing fluctuations are strongly enhanced, as a function of the negative effective range.
As an application to neutron star physics, we also demonstrate how effects of pairing fluctuations in the small-positive-effective-range region can be extracted from results in the negative effective range region.  
\par
This paper is organized as follows. 
In Sec. \ref{sec2}, we present the formalism of the non-selfconsistent $T$-matrix approximation with the coupled fermion-boson model. 
In Sec. \ref{sec3}, we first review the BCS-BEC crossover physics with the negative effective range in this model.
Subsequently, we present numerical results of the single-particle density of states in the BCS-BEC crossover regime.  
Throughout this paper, for simplicity, we set $\hbar=k_{\rm B}=1$ and the system volume is taken to be unity.
\par
\section{Formulation}
\label{sec2}
We start from the coupled fermion-boson model described by the Hamiltonian \cite{Timmermans,Holland,Ohashi2,Liu},
\begin{eqnarray}
\label{eq1}
H&=&\sum_{\bm{p},\sigma}\xi_{\bm{p}}c_{\bm{p},\sigma}^{\dag}c_{\bm{p},\sigma}
+\sum_{\bm{q}}\left(\varepsilon_{\bm{q}}/2+2\nu-2\mu\right)b_{\bm{q}}^{\dag}b_{\bm{q}} \cr
&&+g_{\rm r}\sum_{\bm{p},\bm{q}}\left(b_{\bm{q}}^{\dag}c_{\bm{p}+\bm{q}/2,\up}c_{-\bm{p}+\bm{q}/2,\dwn}+{\rm H.c.}\right).
\end{eqnarray}
Here, $c_{\bm{p},\sigma}$ and $b_{\bm{q}}$ are the annihilation operators of a Fermi atom with the pseudospin $\sigma=\up,\dwn$ and a diatomic molecular boson, respectively. 
$\xi_{\bm{p}}=\varepsilon_{\bm{p}}-\mu$ is the kinetic energy of Fermi atoms measured from the chemical potential $\mu$, where $\varepsilon_{\bm{p}}=p^2/2m$ ($m$ is an atomic mass).
The threshold energy of the diatomic molecule $2\nu$ and the Feshbach coupling constant $g_{\rm r}$ are related to the two-body scattering length $a$ and the effective range $r_{\rm e}$, respectively.
These relations are given by
\begin{eqnarray} 
\label{eq2}
\frac{4\pi a}{m}=-g_{\rm r}^2\left[2\nu-\sum_{\bm{p}}\frac{g_{\rm r}^2}{2\varepsilon_{\bm{p}}}\right]^{-1}\equiv-\frac{g_{\rm r}^2}{2\nu_{\rm r}},
\end{eqnarray}
\begin{eqnarray}
\label{eq3}
r_{\rm e}=-\frac{8\pi}{m^2 g_{\rm r}^2}.
\end{eqnarray}
In Eq. (\ref{eq2}), $2\nu_{\rm r}$ is the renormalized threshold energy. 
For simplicity, we ignore the existence of non-resonant atom-atom scatterings.
\par
\begin{figure}[t]
\begin{center}
\includegraphics[width=6cm]{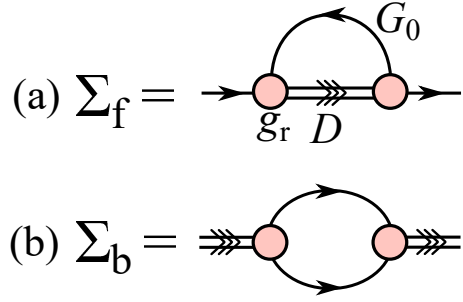}
\end{center}
\caption{The self-energy corrections of (a) Fermi atoms and (b) diatomic molecules.
The single line ($G_0$) and double line ($D$) represent Green's functions of non-interacting atoms and  dressed molecules, respectively.
The shaded circle is the Feshbach coupling $g_{\rm r}$. }
\label{fig1}
\end{figure}
We consider strong coupling effects in the framework of non-selfconsistent $T$-matrix approximation.
The thermal Green's function of a Fermi atom $G$ is given by
\begin{eqnarray}
\label{eq4}
G(\bm{p},i\omega_{n})=\frac{1}{i\omega_n-\xi_{\bm{p}}-\Sigma_{\rm f}(\bm{p},i\omega_{n})},
\end{eqnarray}
where $\omega_n=(2n+1)\pi T$ is the fermion Matsubara frequency.
Figure \ref{fig1}(a) shows the diagrammatic representation of the self-energy $\Sigma_{\rm f}(\bm{p},i\omega_{n})$, which is in the form of,
\begin{eqnarray}
\label{eq5}
\Sigma_{\rm f}(\bm{p},i\omega_{n})=T\sum_{\bm{q},\zeta_{l}}g_{\rm r}^2D(\bm{q},i\zeta_{l})G_0(\bm{q}-\bm{p},i\zeta_l - i\omega_{n}),
\end{eqnarray}
where $G_0(\bm{p},i\omega_{n})=(i\omega_n-\xi_{\bm{p}})^{-1}$ is the bare Green's function of the Fermi atoms and $\zeta_l=2l\pi T$ is the boson Matsubara frequency. 
The thermal Green's function of a dressed molecule $D(\bm{q},i\zeta_{l})$ includes the self-energy correction $\Sigma_{\rm b}(\bm{q},i\zeta_l)$ shown in Fig. \ref{fig1}(b) as follows: 
\begin{eqnarray}
\label{eq6}
D(\bm{q},i\zeta_{l})=\frac{1}{i\zeta_l-\varepsilon_{\bm{q}}/2-2\nu+2\mu-\Sigma_{\rm b}(\bm{q},i\zeta_l)}.
\end{eqnarray}
$\Sigma_{\rm b}(\bm{q},i\zeta_{l})$ is given by,
\begin{eqnarray}
\label{eq7}
\Sigma_{\rm b}(\bm{q},i\zeta_{l})=-g_{\rm r}^2\Pi(\bm{q},i\zeta_{l}),
\end{eqnarray}
where,
\begin{eqnarray}
\label{eq8}
\Pi(\bm{q},i\zeta_{l})&=&T\sum_{\bm{p},i\omega_n}G_0(\bm{p}+\bm{q}/2,i\omega_n+i\zeta_l)G_0(-\bm{p}+\bm{q}/2,-i\zeta_l) \cr
&=&-\sum_{\bm{p}}\frac{1-f(\xi_{\bm{p}+\bm{q}/2})-f(\xi_{-\bm{p}+\bm{q}/2})}{i\zeta_l -\xi_{\bm{p}+\bm{q}/2}-\xi_{-\bm{p}+\bm{q}/2}},         
\end{eqnarray}
is the lowest-order particle-particle correlation function.
In Eq. (\ref{eq8}), $f(x)=1/(e^{x/T}+1)$ is the Fermi-Dirac distribution function.
We note that the ultraviolet divergence of summation of $\bm{p}$ in Eq. (\ref{eq8}) can be avoided by the renormalization of $\nu$.
In this regard, Eq. (\ref{eq6}) can be rewritten as
\begin{eqnarray}
\label{eq9}
D(\bm{q},i\zeta_l)=\frac{1}{i\zeta_l-\varepsilon_{\bm{q}}/2-2\nu_{\rm r}+2\mu+g_{\rm r}^2\left[\Pi(\bm{q},i\zeta_l)-\sum_{\bm{p}}\frac{1}{2\varepsilon_{\bm{p}}}\right]}. 
\end{eqnarray}
\par
The superfluid phase transition temperature $T_{\rm c}$ is determined by the Hugenholtz-Pines condition \cite{Hugenholtz} of Feshbach molecular bosons $[D(\bm{q}=0,i\zeta_l=0)]^{-1}=0$, which reads,
\begin{eqnarray}
\label{eq10}
\frac{m}{4\pi a}+\frac{2\mu}{g_{\rm r}^2}+\sum_{\bm{p}}\left[\frac{1}{2\xi_{\bm{p}}}\tanh\left(\frac{\xi_{\bm{p}}}{2T_{\rm c}}\right)-\frac{1}{2\varepsilon_{\bm{p}}}\right]=0.
\end{eqnarray}
Eq. (\ref{eq10}) is equivalent to the so-called Thouless criterion and recovers the ordinary gap equation of the single-channel model \cite{Tsuchiya} at the broad-resonance limit ($g_{\rm r}\rightarrow \infty$).
We determine $T_{\rm c}$ and the critical chemical potential $\mu_{\rm c}=\mu(T=T_{\rm c})$ by self-consistently solving Eq. (\ref{eq10}) and particle number equation,
\begin{eqnarray}
\label{eq11}
N&=&2N_{\rm f}+2N_{\rm b}\cr
&=&2T\sum_{\bm{p},i\omega_n}G(\bm{p},i\omega_n)+2T\sum_{\bm{q},i\zeta_l}D(\bm{q},i\zeta_l),
\end{eqnarray}
where $N$ is the total number.
$N_{\rm f}$ and $N_{\rm b}$ are the numbers of Fermi atoms and Feshbach molecules, respectively.
\par
In this paper, we calculate the single-particle density of states of a Fermi atom given by
\begin{eqnarray}
\label{eq12}
\rho(\omega)&=&\sum_{\bm{p}}A(\bm{p},\omega) \cr
&=&-\frac{1}{\pi}\sum_{\bm{p}}{\rm Im}G(\bm{p},i\omega_n\rightarrow \omega+i\delta),
\end{eqnarray}
where $A(\bm{p},\omega)$ is the single-particle spectral function and $\omega$ is the single-particle energy. 
In Eq. (\ref{eq12}), the analytic continuation ($i\omega_n\rightarrow \omega+i\delta$) is numerically done by using the Pad\'{e} approximation with the small number $\delta=10^{-2}\varepsilon_{\rm F}$, where $\varepsilon_{\rm F}$ is the Fermi energy.
\par
\section{Results}
\label{sec3} 
\begin{figure}[t]
\begin{center}
\includegraphics[width=6.5cm]{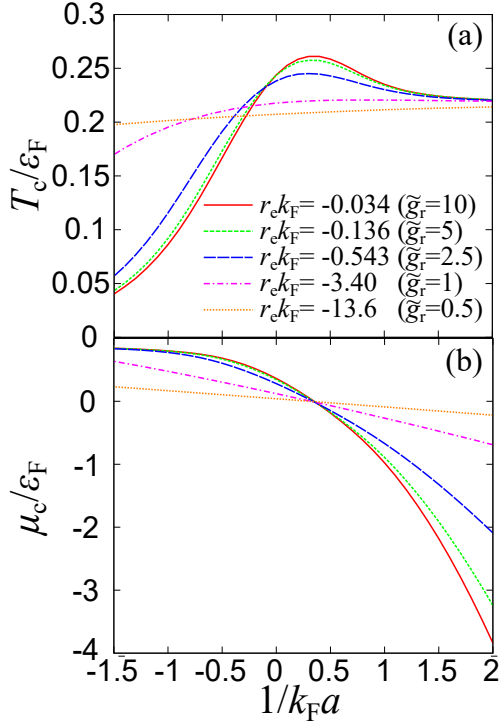}
\end{center}
\caption{(a) The superfluid phase transition temperature $T_{\rm c}$ and (b) the critical chemical potential $\mu_{\rm c}$ in the whole BCS-BEC crossover regime with the finite effective range $r_{\rm e}$.
The quantity $\tilde{g}_{\rm r}=g_{\rm r}\sqrt{N}/\varepsilon_{\rm F}$ is the dimensionless Feshbach coupling.}
\label{fig2}
\end{figure}

At first, we show the effective-range (Feshbach coupling) dependence of the superfluid phase transition temperature $T_{\rm c}$ and the critical chemical potential $\mu_{\rm c}$ in the BCS-BEC crossover regime in Fig. \ref{fig2}.
Here, $\tilde{g}_{\rm r}=g_{\rm r}\sqrt{N}/\varepsilon_{\rm F}$ is the dimensionless Feshbach coupling, which is connected with the scaled effective range $r_{\rm e}k_{\rm F}=-32/(3\pi \tilde{g}_{\rm r}^2)$.
In the broad-resonance regime ($|r_{\rm e}k_{\rm F}|\lesssim 1$), $T_{\rm c}$ and $\mu_{\rm c}$ are almost equal to the results of a previous work on the single-channel model \cite{Tsuchiya}.
In the strong-coupling BEC regime ($1/k_{\rm F}a$\gesim 1), $T_{\rm c}$ and $\mu_{\rm c}$ go to the BEC temperature of tightly bound molecular bosons $T_{\rm c}^{\rm BEC}=0.218\varepsilon_{\rm F}$ and the half of their binding energy $E_{\rm b}/2=-1/(2ma^2)$, respectively \cite{Nozieres, SadeMelo,Tsuchiya}.
On the other hand, in the weak-coupling BCS regime ($1/k_{\rm F}a$\lesssim -1), $T_{\rm c}$ approaches the famous BCS superfluid phase transition temperature $T_{\rm c}^{\rm BCS}\simeq0.614T_{\rm F}e^{-\frac{\pi}{2k_{\rm F}a}}$ and $\mu_{\rm c}$ becomes close to $\varepsilon_{\rm F}$ \cite{Leggett}. 
\par
In the narrow-resonance regime, or large-negative-effective-range region ($|r_{\rm e}k_{\rm F}|\gesim 1$),
$T_{\rm c}$ and $\mu_{\rm c}$ deviate from results in the broad-resonance region. 
$T_{\rm c}$ increases with decreasing $\tilde{g}_{\rm r}$ where $\mu_{\rm c}$ is positive ($1/k_{\rm F}a \lesssim 0.31$).
This enhancement of $T_{\rm c}$ is consistent with the previous work \cite{Ho2} which suggests that the narrow Feshbach resonance produces strong-pairing effects where the two-body bound state is absent.
We note that $T_{\rm c}$ slightly decreases in the opposite side ($1/k_{\rm F}a \gesim 0.31$).
$\mu_{\rm c}$ approaches $0$ in the whole crossover region with decreasing $g_{\rm r}$.
In the large-negative-effective-range limit $r_{\rm e}\rightarrow -\infty$ (narrow-resonance limit $g_{\rm r}\rightarrow 0$), 
the system can be exactly described by the mean-field theory since the self-energy corrections given by Eqs. (\ref{eq5}) and (\ref{eq7}) are proportional to $g_{\rm r}^2$.
In this case, Eq. (\ref{eq11}) becomes
\begin{eqnarray}
\label{eq13}
N&=&2N_{\rm f}^0+2N_{\rm b}^0 \cr
&=&2\sum_{\bm{p}}f\left(\xi_{\bm{p}}\right)+2\sum_{\bm{q}}b\left(\varepsilon_{\bm{q}}/2+2\nu_{\rm r}-2\mu\right),
\end{eqnarray}
where $b(x)=1/(e^{x/T}-1)$ is the Bose-Einstein distribution function.
$N_{\rm f}^0$ and $N_{\rm b}^0$ in Eq. (\ref{eq13}) represent the particle number of non-interacting Fermi atoms and diatomic molecules, respectively.
One can evaluate $\mu_{\rm c}$ from Eq. (\ref{eq13}) with the condition of the gapless bosonic excitation as,
\begin{eqnarray}
\label{eq14}
\mu_{\rm c}=\nu_{\rm r}=-\frac{mg_{\rm r}^2}{8\pi a},
\end{eqnarray}
which indicates $\mu_{\rm c}=0$ in this limit ($g_{\rm r}\rightarrow 0$) with the finite scattering length.
Substituting $\mu_{\rm c}=0$ to Eq. (\ref{eq13}), one can obtain the critical temperature of the narrow resonance limit,
\begin{equation}
\label{eq15}
T_{\rm c}^{\rm NRL}=0.204T_{\rm F}.
\end{equation}
Actually, $T_{\rm c}$ at $r_{\rm e}k_{\rm F}=-13.6$ ($\tilde{g}_{\rm r}=0.5$) shown in Fig. \ref{fig2} (a) is very close to $T_{\rm c}^{\rm NRL}$.    
\par
\begin{figure}[t]
\begin{center}
\includegraphics[width=6.5cm]{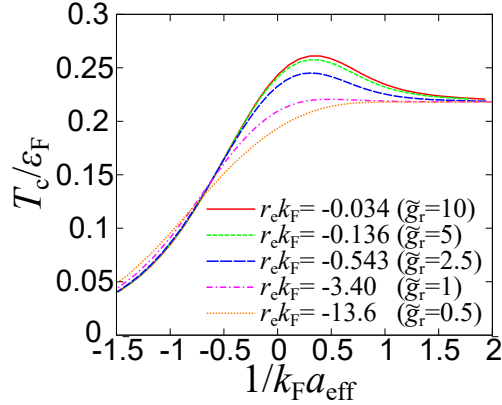}
\end{center}
\caption{The superfluid phase transition temperature $T_{\rm c}$ at various Feshbach couplings, as a function of $1/k_{\rm F}a_{\rm eff}$ where $a_{\rm eff}$ is the effective scattering length defined by Eq. (\ref{eq16}).}
\label{fig3}
\end{figure}
In the narrow-resonance regime, it is known that an effective scattering length $a_{\rm eff}$ \cite{Liu,Ohashi7} is useful to measure the interaction strength in the BCS-BEC crossover regime, defined by
\begin{eqnarray}
\label{eq16}
\frac{4\pi a_{\rm eff}}{m}=-\frac{g_{\rm r}^2}{2\nu_{\rm r}-2\mu},
\end{eqnarray}
since the effective-interaction strength between Fermi atoms is given by $U_{\rm eff}(\bm{q},i\zeta_l)=-g_{\rm r}^2D(\bm{q},i\zeta_l)$.
Indeed, using $a_{\rm eff}$, one can find that Eq. (\ref{eq10}) can be rewritten as
\begin{eqnarray}
\label{eq17}
\frac{m}{4\pi a_{\rm eff}}+\sum_{\bm{p}}\left[\frac{1}{2\xi_{\bm{p}}}\tanh\left(\frac{\xi_{\bm{p}}}{2T_{\rm c}}\right)-\frac{1}{2\varepsilon_{\bm{p}}}\right]=0.
\end{eqnarray}
Eq. (\ref{eq17}) is the same form of the ordinary BCS gap equation in the single-channel model \cite{Tsuchiya}.
Fig. \ref{fig3} shows $T_{\rm c}$ as a function of $1/k_{\rm F}a_{\rm eff}$ at various Feshbach couplings.
Although $T_{\rm c}$ quantitatively changes if one tunes $g_{\rm r}$ in the intermediate region ($-1\lesssim1/k_{\rm F}a_{\rm eff}\lesssim 1$), both weak-coupling and strong-coupling regimes except the above region do not depend on $g_{\rm r}$.
In this regard, in the case of narrow resonance, it is appropriate that the weak-coupling BCS regime and the strong-coupling BEC regime are defined as $1/k_{\rm F}a_{\rm eff}\lesssim-1$ and $1/k_{\rm F}a_{\rm eff}\gesim 1$, respectively.
\begin{figure}[t]
\begin{center}
\includegraphics[width=6.5cm]{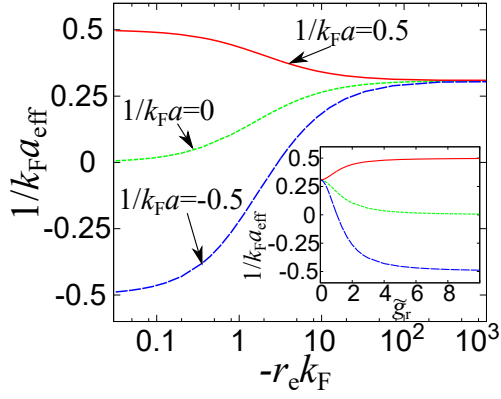}
\end{center}
\caption{The effective range dependence of the inverse effective scattering length $1/k_{\rm F}a_{\rm eff}$ at $1/k_{\rm F}a=-0.5$ (solid line), $0$ (dotted line) and $0.5$ (dashed line) at $T=T_{\rm c}$.
The inset shows $1/k_{\rm F}a_{\rm eff}$ as a function of the dimensionless Feshbach coupling $\tilde{g}_{\rm r}$.
In each figure, we use the same line style at each scattering length.}
\label{fig4}
\end{figure}
\par
Fig. \ref{fig4} shows the effective-range dependence of $1/k_{\rm F}a_{\rm eff}$ in the crossover region ($1/k_{\rm F}a=-0.5, 0$, and $0.5$) at $T=T_{\rm c}$.
We also show the Feshbach coupling dependence of $1/k_{\rm F}a_{\rm eff}$ in the inset of Fig. \ref{fig4}.
In the narrow-resonance limit, one can find that $1/k_{\rm F}a_{\rm eff}\simeq0.31$ (where $\mu_{\rm c}=0$) at each scattering length. 
This is nothing but the reason why the narrow Feshbach resonance induces a strong attraction between Fermi atoms and $T_{\rm c}$ is enhanced by the negative effective range in the region where $1/k_{\rm F}a\lesssim 0$ shown in Fig. \ref{fig2} (a).
\begin{figure}[t]
\begin{center}
\includegraphics[width=6cm]{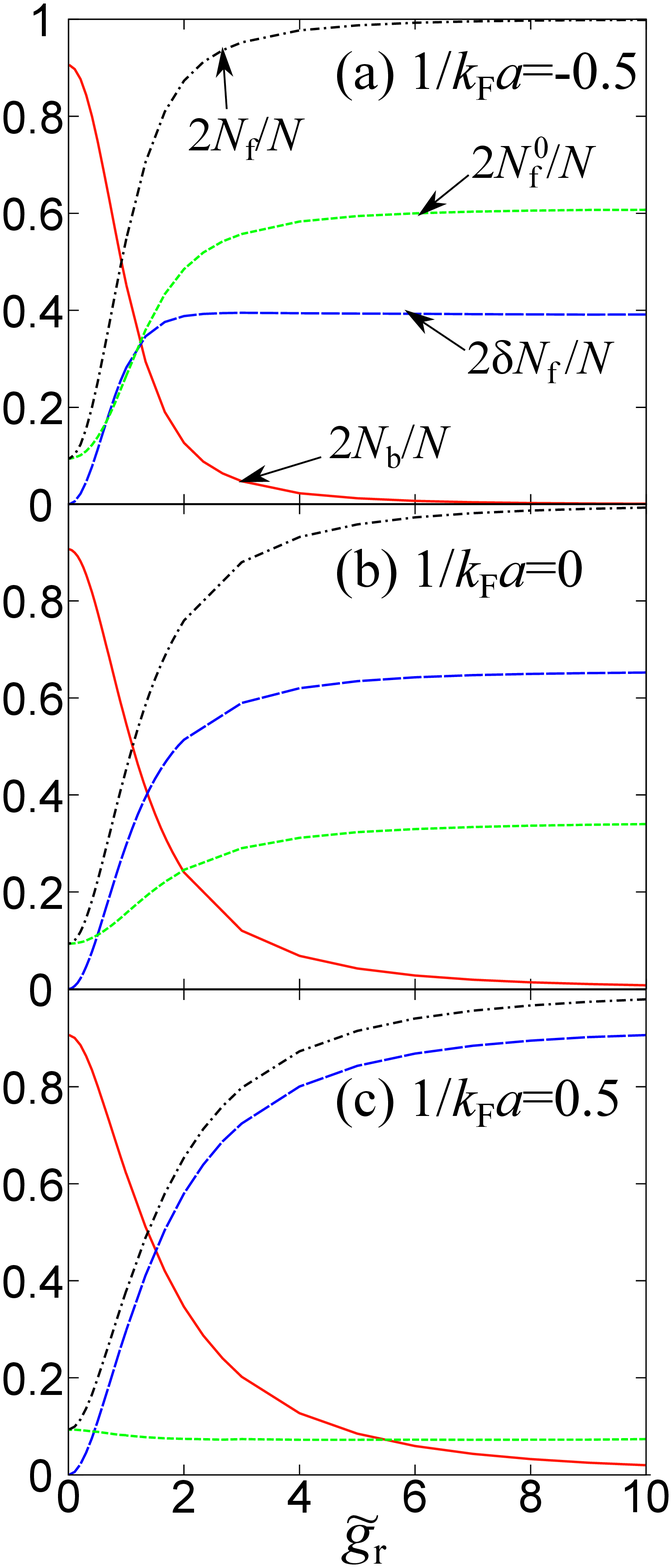}
\end{center}
\caption{Particle numbers as a function of dimensionless Feshbach coupling $\tilde{g}_{\rm}$ at $1/k_{\rm F}a=-0.5$ (a), $0$ (b), and $0.5$ (c) at $T=T_{\rm c}$.}
\label{fig5}
\end{figure}
\par
In Fig. \ref{fig5}, we show the Feshbach coupling dependence of particle numbers in the crossover region.
The particle number of Fermi atoms $2N_{\rm f}$ are divided into two parts,
\begin{eqnarray}
\label{eq18}
2N_{\rm f}&=&2N_{\rm F}^0+2\delta N_{\rm f} \cr
&=&2\sum_{\bm{p}}f(\xi_{\bm{p}})+2T\sum_{\bm{p},i\omega_n}\left[G(\bm{p},i\omega_n)-G_0(\bm{p},i\omega_n)\right],          
\end{eqnarray}
where the second term $2\delta N_{\rm f}$ is the fluctuation corrections.
$\delta N_{\rm f}$ monotonically increases with increasing the interaction strength from $1/k_{\rm F}a=-0.5$ [Fig. \ref{fig5}(a)] to $0.5$ [Fig. \ref{fig5}](c).
On the other hand, $2\delta N_{\rm f}$ monotonically decreases with decreasing $g_{\rm r}$ at each scattering length and the total Fermi atomic number becomes dominated by the non-interacting part $2N_{\rm f}^0$.
In the narrow resonance limit, $2N_{\rm f}^0$ at $T=T_{\rm c}^{\rm NRL}$ ($\mu_{\rm c}=0$) approaches a constant value given by
\begin{eqnarray}
\label{eq19}
2N_{\rm f}&\simeq&2N_{\rm f}^0=2\sum_{\bm{p}}f(\varepsilon_{\bm{p}}) \cr
&\simeq&0.0937N.
\end{eqnarray}
In contrast to $2N_{\rm f}$, the particle number of diatomic molecular bosons $2N_{\rm b}$ increases with decreasing $g_{\rm r}$ and finally reaches $N-2N_{\rm f}^0\simeq 0.906N$ in the narrow resonance limit. 
This interplay of $2N_{\rm f}$ and $2N_{\rm b}$ and the suppression of the fluctuation contribution $2\delta N_{\rm f}$ in spite of the strong attraction between atoms are characteristic features of narrow Feshbach resonances that can not be seen in broad Feshbach resonances.
\par
\begin{figure}[t]
\begin{center}
\includegraphics[width=6.5cm]{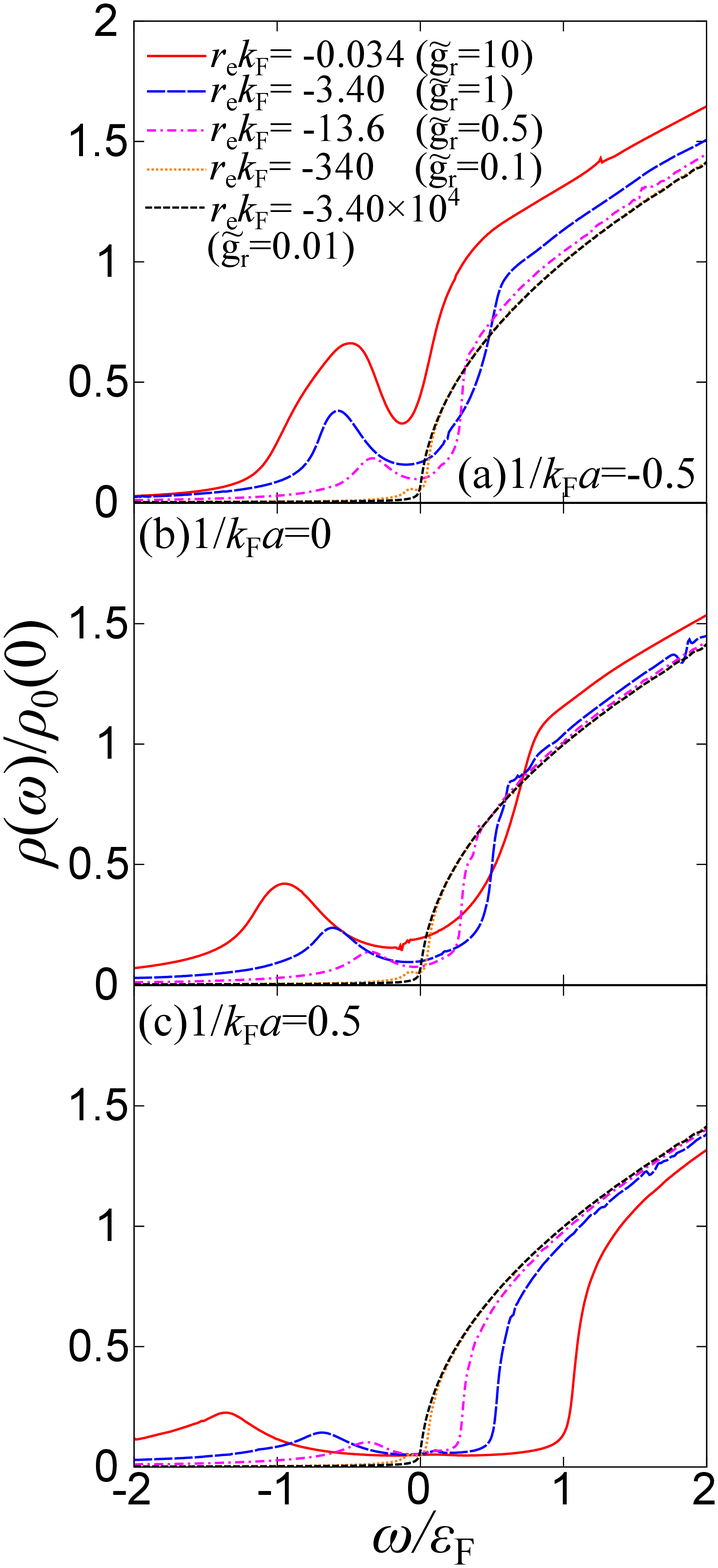}
\end{center}
\caption{Single-particle density of states $\rho(\omega)$ at $T=T_{\rm c}$ with various effective ranges [$1/k_{\rm F}a=-0.5$ (a), $0$ (b), $0.5$ (c)].
$\rho_0(0)$ is the single-particle density of states at the Fermi level in an ideal Fermi gas at $T=0$.}
\label{fig6}
\end{figure}
Fig. \ref{fig6} shows the single-particle density of states $\rho(\omega)$ in the crossover regime at $T=T_{\rm c}$ with negative effective range, where $\rho_0(\omega=0)=mk_{\rm F}/(2\pi^2)$ is the single-particle density of states at the Fermi level in a non-interacting Fermi gas at $T=0$.
In the weak-coupling side $1/k_{\rm F}a=-0.5$ (a), 
one can see that the pseudogap phenomenon appears as a dip structure around $\omega=0$ at the broad resonance ($r_{\rm e}k_{\rm F}=-0.034$).
This pseudogap size is enhanced with decreasing $g_{\rm r}$ in the broad resonance region ($\tilde{g}_{\rm r}\gesim 1$).
However, the pseudogap closes with decreasing $g_{\rm r}$ in the narrow resonance region ($\tilde{g}_{\rm r}\lesssim 1$).
This fact can be understood by considering the static approximation \cite{Tsuchiya,Perali} given by
\begin{eqnarray}
\label{eq20}
\Sigma_{\rm f}(\bm{p},i\omega_n)&\simeq&-\Delta_{\rm pg}^2G_0(-\bm{p},-i\omega_n),
\end{eqnarray}
where,
\begin{eqnarray}    
\label{eq21}
\Delta_{\rm pg}^2 &=&-T\sum_{\bm{q},i\zeta_l}g_{\rm r}^2D(\bm{q},i\zeta_l) \cr
&=&g_{\rm r}^2N_{\rm b},
\end{eqnarray}
is the so-called pseudogap parameter which is directly related to $g_{\rm r}$ as well as $N_{\rm b}$.
We note that this approximation is justified near $T_{\rm c}$ where $D(\bm{q}=0,i\zeta_l=0)$ diverges.
By substituting Eq. (\ref{eq20}) to Eq. (\ref{eq4}), one can obtain,
\begin{eqnarray}
\label{eq22}
G(\bm{p},i\omega_n)\simeq\frac{i\omega_n+\xi_{\rm p}}{(i\omega_n)^2 - \xi_{\bm{p}}^2-\Delta_{\rm pg}^2}.
\end{eqnarray}
Eq. (\ref{eq22}) shows that $G(\bm{p},i\omega_n)$ becomes similar to the BCS Green's function even above $T_{\rm c}$ due to strong pairing fluctuations.
In this regard, the pseudogap size is determined by $\Delta_{\rm pg}$ where $\mu>0$.
Since $N_{\rm b}$ monotonically increases with decreasing $g_{\rm r}$ as shown in Fig. \ref{fig5}, 
$\Delta_{\rm pg}$ also increases in the broad-resonance region.
However, in the narrow-resonance region, $\Delta_{\rm pg}$ is proportional to $g_{\rm r}$ and disappear at $g_{\rm r}\rightarrow 0$ because $N_{\rm b}$ becomes almost constant.
On the other hand, in the strong-coupling side $1/k_{\rm F}a=0.5$ (c) where $\mu_{\rm c}<0$,
the gap size in $\rho(\omega)$ is given by $2\sqrt{\mu_{\rm c}^2+\Delta_{\rm pg}^2}$.
This value depends on $\mu_{\rm c}$ rather than $\Delta_{\rm pg}$.
In the strong-coupling limit with the broad resonance, 
this energy gap is given by the binding energy of bound molecules where $2|\mu_{\rm c}|\simeq E_{\rm b}=1/ma^2$ \cite{Leggett,Tsuchiya}.
In the narrow-resonance regime where $\mu_{\rm c}<0$, the energy gap monotonically disappears since both $\mu_{\rm c}$ and $\Delta_{\rm pg}$ approach $0$ with decreasing $g_{\rm r}$.
\par
We note that in the case of the broad-resonance limit ($r_{\rm e}\rightarrow 0$), $\Delta_{\rm pg}$ is related to Tan's contact $C$ \cite{Tan1,Tan2,Tan3}, which can be represented by $C=m^2\Delta_{\rm pg}^2$ \cite{Palestini2}.
In the coupled fermion-boson model, it is given by \cite{Kamikado},
\begin{eqnarray}
\label{eq23} 
C=m^2g_{\rm r}^2N_{\rm b}=-\frac{8\pi N_{\rm b}}{r_{\rm e}}.
\end{eqnarray}
Although Tan's relation is developed in the case of the zero-range contact potential, 
how the finite effective range affects $C$ in the whole BCS-BEC crossover region is an interesting problem left as a future work.
\par
\begin{figure}[t]
\begin{center}
\includegraphics[width=6.5cm]{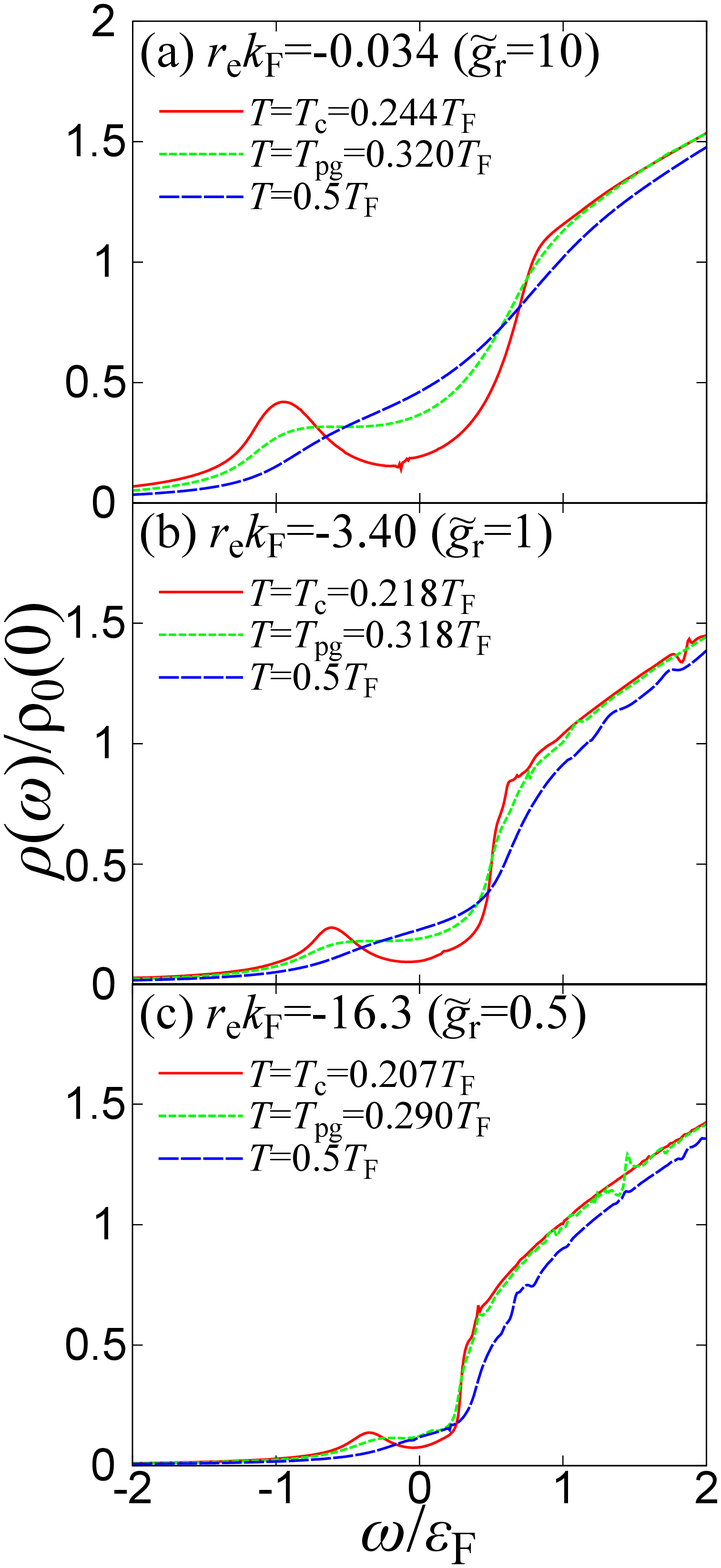}
\end{center}
\caption{The single-particle density of states $\rho(\omega)$ at $1/k_{\rm F}a=0$ at the different temperatures.
In each figure, the solid, dotted, and dashed lines represent results of $T=T_{\rm c}$, $T_{\rm pg}$, and $0.5T_{\rm F}$, respectively.
Here $T_{\rm pg}$ is the pseudogap temperature defined as the temperature where the dip structure in $\rho(\omega)$ disappears.
We set the effective range $r_{\rm e}k_{\rm F}=-0.034$ (a), $-3.40$ (b), and $-16.3$ (c).
}
\label{fig7}
\end{figure}
Figure \ref{fig7} shows the temperature dependence of $\rho(\omega)$ at $1/k_{\rm F}a=0$, 
where $r_{\rm e}k_{\rm F}=-0.034$ (a), $-3.40$ (b) and $-16.3$ (c).
The pseudogap is gradually smeared due to thermal fluctuations with increasing $T$ and the dip structure disappears at high temperature.
In the high temperature region ($T\gesim 0.5T_{\rm F}$), $\rho(\omega)$ qualitatively corresponds to the density of states in a non-interacting Fermi gas given by $\rho_0(\omega)=\frac{m}{2\pi^2}\sqrt{2m(\omega+\mu)}$.
In the narrow-resonance region [$r_{\rm e}k_{\rm F}=-16.3$ as shown in Fig. \ref{fig7} (c)], 
the pseudogap structure disappears at lower temperature than in the case of the broad Feshbach resonance since the pseudogap size is also smaller at $T=T_{\rm c}$.
In this paper, we introduce the pseudogap temperature $T_{\rm pg}$ \cite{Tsuchiya} which is a characteristic temperature where the dip structure in $\rho(\omega\simeq 0)$ disappears, shown as the dotted lines in Fig. \ref{fig7}.
Although the definition of this characteristic temperature has some  ambiguity because the pseudogap is a crossover phenomenon without any distinct changes of properties like a phase transition, one can expect that the system properties are dominated by strong pairing fluctuations which cannot be explained by the mean-field theory or the Fermi-liquid theory below $T_{\rm pg}$.
Actually, a similar characteristic temperature can be observed via the temperature dependence of thermodynamic quantities such as specific heat \cite{Pieter2} and spin susceptibility \cite{Tajima}.
\par     
\begin{figure}[t]
\begin{center}
\includegraphics[width=6.5cm]{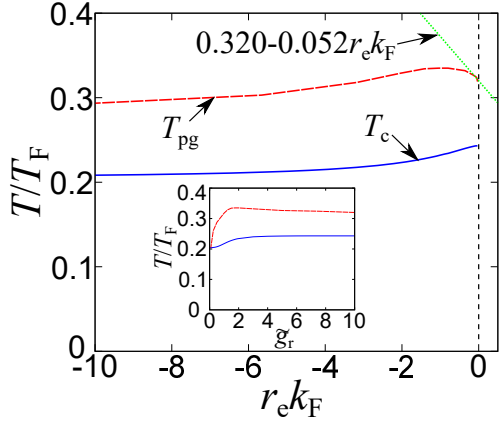}
\end{center}
\caption{The superfluid phase transition temperature $T_{\rm c}$ (solid line) and the pseudogap temperature $T_{\rm pg}$ (dashed line) as a function of the effective range $r_{\rm e}$ at $1/k_{\rm F}a=0$.
The dotted line shows the linear fitting of $T_{\rm pg}$ in the small negative effective range region ($|r_{\rm e}k_{\rm F}|\lesssim 0.05$).
The inset shows $T_{\rm c}$ and $T_{\rm pg}$ versus the dimensionless Feshbach coupling $\tilde{g}_{\rm r}$.
}
\label{fig8}
\end{figure}
Fig. \ref{fig8} shows the negative-effective-range dependence of $T_{\rm c}$ and $T_{\rm pg}$ at $1/k_{\rm F}a=0$.
In the broad-resonance region ($r_{\rm e}k_{\rm F}\gesim-1$), $T_{\rm pg}$ is slightly enhanced with decreasing $r_{\rm e}$ reflecting the increase of $\Delta_{\rm pg}$.
In the narrow resonance regime ($r_{\rm e}k_{\rm F}\lesssim -1$), $T_{\rm pg}$ gradually decreases with decreasing $r_{\rm e}$ and as shown in the inset of Fig. \ref{fig8}, $T_{\rm pg}$ coincides with $T_{\rm c}$ around $\tilde{g}_{\rm r}\simeq0.08$, where the corresponding effective range is given by $r_{\rm e}k_{\rm F}\simeq-5.3\times 10^2$.  
Beyond this value, the system properties can be described by the mean-field theory even near $T_{\rm c}$.
\par
Using results shown in Fig. \ref{fig8}, we demonstrate applications to dilute neutron matter which has a positive effective range in the neutron-neutron scattering.
At $1/k_{\rm F}a=0$, it is known that the ground-state energy $E(r_{\rm e}k_{\rm F})$ with a small effective range can be expressed as \cite{Carlson,Schonenberg},
\begin{eqnarray}
\label{eq25}
\frac{E(r_{\rm e}k_{\rm F})}{E_{\rm FG}}=\xi_{\rm B}+\zeta r_{\rm e}k_{\rm F} + O(r_{\rm e}^2k_{\rm F}^2),
\end{eqnarray}
where $E_{\rm FG}=\frac{3}{5}N\varepsilon_{\rm F}$ is the ground-state energy of an ideal Fermi gas.
In Eq. (\ref{eq25}), $\xi_{\rm B}$ and $\zeta$ are the Bertsch parameter \cite{Baker} and the linear coefficient with respect to $r_{\rm e}k_{\rm F}$, respectively.
Recently, $\xi_{\rm B}$ and $\zeta$ have been determined by QMC simulations \cite{Forbes,Carlson,Schonenberg}.
Moreover, $\xi_{\rm B}$ has been precisely measured in current experiments \cite{Ku,Horikoshi,Zurn}.
Analogously, we expand $T_{\rm pg}$ with respect to $r_{\rm e}k_{\rm F}$ and determine the linear coefficient from the fitting of $T_{\rm pg}$ in the small-negative-effective-range region ($|r_{\rm e}k_{\rm F}|\lesssim 0.05$). 
As a result, we obtain
\begin{eqnarray}
\label{eq26}
\frac{T_{\rm pg}(r_{\rm e}k_{\rm F})}{T_{\rm F}}=0.320-0.052r_{\rm e}k_{\rm F} + O(r_{\rm e}^2k_{\rm F}^2).
\end{eqnarray}
In the small effective range region, one can expect that Eq. (\ref{eq26}) is valid even for the {\it positive} effective range.
In this sense, from Eq. (\ref{eq26}), one can find that pairing fluctuations seem to be suppressed by the positive effective range since $T_{\rm pg}$ decreases with increasing $r_{\rm e} (>0)$.
This estimation is expected to be reasonable since the positive effective range suppresses the magnitude of the scattering phase shift which characterizes the interaction strength.
It can be an important information for astrophysical simulations or studies on the cooling process of a neutron star \cite{Oertel}.
We emphasize that this characteristic temperature originating from strong pairing fluctuations can be determined in cold-atom experiments through the measurement of thermodynamic quantities such as spin susceptibility, which is now experimentally accessible \cite{Sanner,Sommer,Meineke}. 
\par
We note that the same analysis can be applied to $T_{\rm c}$ but it is necessary to consider effects of particle-hole fluctuations \cite{Pisani,Floerchinger,Yu} to obtain the correct effective-range dependence of $T_{\rm c}$.
Indeed, the non-selfconsistent $T$-matrix approximation overestimates $T_{\rm c}\simeq 0.24T_{\rm F}$ in the unitarity limit ($1/k_{\rm F}a=0$, $r_{\rm e}=0$) compared to the experimental value $0.167(13)T_{\rm F}$ \cite{Ku}.  
Although the particle-hole fluctuations may affect $T_{\rm pg}$, we expect that our result for $T_{\rm pg}$ is qualitatively unchanged since the non-selfconsistent $T$-matrix approximation can successfully explain effects of pairing fluctuations on the recent experimental results of local photoemission spectra in the pseudogap regime \cite{Ota}.

\section{Summary}
\label{sec4}
To summarize, we have theoretically investigated the effects of pairing fluctuations in a strongly interacting Fermi gas with negative effective range.
Within the framework of the non-selfconsistent $T$-matrix approximation with the coupled fermion-boson model for the narrow Feshbach resonance, 
we have discussed the negative-effective-range corrections on the single-particle density of states at the superfluid phase transition temperature $T_{\rm c}$ in the BCS-BEC crossover regime.
\par
On the weak coupling side $1/k_{\rm F}a\lesssim 0$ where the critical chemical potential is positive ($\mu_{\rm c}>0$), the negative-effective-range corrections induce strong pairing effects and the pseudogap size at $T_{\rm c}$ is enhanced in the broad-resonance regime ($|r_{\rm e}k_{\rm F}|\lesssim 1$).
On the other hand, on the strong coupling side $1/k_{\rm F}a\gesim 0$ where $\mu_{\rm c}<0$, the effective interaction strength is weakened due to the presence of the negative effective range and the pseudogap size monotonically decreases with decreasing $r_{\rm e}$.
Approaching the narrow-resonance limit ($r_{\rm e}\rightarrow -\infty$, $g_{\rm r}\rightarrow 0$), the system's properties are exactly described by the mean field theory and the pseudogap disappears at each scattering length.
\par
At $1/k_{\rm F}a=0$, we have shown the negative-effective-range dependence of the pseudogap temperature $T_{\rm pg}$, which is one of the characteristic temperatures where strong pairing fluctuations affect physical quantities.
While in the broad-resonance region ($|r_{\rm e}k_{\rm F}|\lesssim 1$), $T_{\rm pg}$ increases with decreasing $r_{\rm pg}$, the pseudogap region ($T_{\rm c}<T<T_{\rm pg}$) disappears in the deep-narrow-resonance regime ($r_{\rm e}k_{\rm F}\lesssim -5.3\times 10^2$).
\par
From the negative-effective-range dependence of $T_{\rm pg}$ in the broad-resonance region, we have obtained $T_{\rm pg}/T_{\rm F}=0.320-0.052r_{\rm e}k_{\rm F}+O(r_{\rm e}^2k_{\rm F}^2)$.
This equation is expected to be valid even in the small-{\it positive}-effective-range region, indicating that pairing fluctuations is suppressed by the small positive effective range.
Since the effects of strong pairing fluctuations near $T_{\rm c}$ in interacting fermions is quite non-trivial and crucial for an ultracold Fermi gas as well as neutron star physics,
our strategy suggests that the experimental realization of a strongly interacting Fermi gas with negative effective range can contribute toward the further understanding of such interdisciplinary topics.
\par
The negative-effective-range dependence of other physical observables remains as an interesting topic for future work.
The diagrammatic approach presented in this paper can be extended to study thermodynamic quantities such as spin susceptibility.
It would also be interesting to study how the negative-effective-range region connects to the positive side in the BCS-BEC crossover regime, as well as the extension to the superfluid phase.   
 
\acknowledgements
The author thanks P. Naidon for kindly reading the manuscript and suggesting improvements,
and T. Hatsuda, Y. Ohashi, S. Uchino, Y. Nishida, and D. Kagamihara for useful discussions.
This work was supported by a Grant-in-Aid for JSPS fellows (No.17J03975).
\par


\end{document}